\documentclass[12pt,preprint]{aastex}

\shorttitle{Magnetic field evolution}
\shortauthors{Wolf et al.}

\begin{document}


\title{Magnetic field evolution in Bok globules}


\author{Sebastian Wolf}
\affil{California Institute of Technology, 1201 East California Blvd, Mail code 105-24, Pasadena, CA 91125, USA; swolf@astro.caltech.edu}

\author{R. Launhardt and Th. Henning}
\affil{Max-Planck-Institut f\"ur Astronomie, K\"onigstuhl 17, D-69117 Heidelberg, Germany; rl@mpia-hd.mpg.de, henning@mpia-hd.mpg.de}




\begin{abstract}
Using the Submillimeter Common-User Bolometer Array (SCUBA)
at the James Clerk Maxwell Telescope (JCMT), we obtained submillimeter
polarization maps of the Bok globules B~335, CB~230, and CB~244 at 850~$\mu$m.
 We find strongly aligned polarization vectors in the case of B~335 and CB~230,
indicating a strong coupling of the magnetic field to the dust grains.
Based on the distribution of the orientation and strength of the linear
polarization we derive the magnetic field strengths in the envelopes
of the globules: 
134$\mu$G (B\,335),
218$\mu$G (CB\,230),
and
257$\mu$G (CB\,244). 
 In agreement with previous submillimeter polarization measurements
of Bok globules we find polarization degrees of several percent
decreasing towards the centers of the cores.
 Furthermore, we compare the magnetic field topology with the spatial structure
of the globules, in particular with the orientation of the outflows and the orientation
of the nonspherical globule cores. In case of the globules B~335 and CB~230, the outflows
are oriented almost perpendicular to the symmetry axis of the globule cores.
The magnetic field, however, is aligned with the symmetry axis
of the prolate cores in the case of the Bok globules B~335 and CB~230, while it
is slightly aligned with the outflow axis in the case of the Bok globules
CB~26 and CB~54. We discuss the possibility that the different orientations
of the magnetic field relative to the outflow directions reflect different
evolutionary stages of the single globules.
\end{abstract}



\section{Introduction}\label{intro}

Bok globules are excellent objects to study the earliest processes
of star formation: They are small in diameter (0.1--2\,pc),
simply structured, and are relatively isolated molecular clouds
(Clemens et al.~1991) with masses of 2--100\,$M_{\rm sun}$ 
(Bok~1977, Leung~1985). Low-mass star formation was found to be a common
phenomenon in Bok globules: 
Many globules have bipolar molecular outflows
(e.g., Yun \& Clemens 1994a) and infrared colors and sub-millimeter
properties that are consistent with Class 0 protostars or embedded
Class I sources Yun \& Clemens 1994b, Launhardt \& Henning 1997 (hereafter LH97);
Henning \& Launhardt 1998).

To study one of the key parameters of the star formation process
-- namely the magnetic field -- submillimeter polarization measurements
represent a powerful technique.
Assuming emission by aligned nonspherical grains as the dominating
polarization mechanism, where the magnetic field plays a role in the alignment process,
magnetic field strengths and structures can be derived from the submillimeter
polarization pattern. In the case of Bok globules this has been demonstrated
for the first time by Henning et al.~(2001; hereafter Paper\,I).
Based on comprehensive preparatory studies such as
submillimeter continuum and CS line surveys (LH97;
Henning \& Launhardt 1998; Launhardt et al. 1997; 1998; in prep.),
polarization maps of three Bok globule cores (CB~26, CB~54, and CG~30)
had been obtained with SCUBA at 850\,$\mu$m.
It was found that the magnetic field strengths derived from polarization patterns 
are well above those of the interstellar medium (Myers et al.~1995),
but are similar to those found in other molecular cloud cores and protostellar
envelopes 
(Bhatt \& Jain 1992;
Levin et al.~2001;
Davis et al.~2000;
Glenn, Walker \& Young 1999;
Itoh et al.\ 1999;
Minchin \& Murray 1994;
Chrysostomou et al.\ 1994;
Crutcher 1999). 
The polarization pattern itself revealed striking similarities
between the different globules: The degree of polarization amounts to several percent and
decreases towards the centers of the dense cores.

Here, we present and discuss new polarization measurements at
850\,$\mu$m of protostellar cores in three other nearby Bok globules:
B\,335 (CB\,199), CB\,230, and CB\,244.
As for the previously observed Bok globules, it was our aim to prove basic correlations 
between the structure and strength of the magnetic field and 
the dust density distribution which have been investigated theoretically, 
e.g., by Padoan et al.~2001, Heitsch et al.~2001, Fiege \& Pudritz~(2000a,c), and Basu \& Mouschovias (1995).
Combining intensity and polarization
maps, the dust density distribution and the magnetic field structure can be found. 
Due to this enlargement of the sample of spatially resolved
polarization maps of Bok globules, the observations were aimed to contribute 
to the solution of the following problems:
\begin{enumerate}
\item Are there systematic differences in the structure and strength
        of the magnetic field in the envelopes around low-mass YSOs
        of different evolutionary stages?
\item Do we see evidence that the magnetic field dominates the structure of globules 
  At which stage of the evolution does the gas decouple from the magnetic fields?
\end{enumerate}

In \S\ref{sources} we compile the main results of previous
investigations of our sources. In \S\ref{obs} we give a brief
overview about the performed observations and the subsequent data reduction
procedure. 
The polarization maps are presented in \S\ref{polpat}; magnetic field
strengths are derived in \S\ref{magfield}, followed by a discussion on the relation 
between the polarization degree and the intensity in \S\ref{plvsi}.
Finally, we investigate the correlation between the magnetic field
structure and morphological features of the individual globules
in \S\ref{magmorph}.

\section{Source description}\label{sources}

In Paper~I we investigated submillimeter polarization maps of the Bok globules CB~26, CB~54,
and CG~30. In this paper we present new polarization measurements of the Bok globules B~335,
(CB~199), CB~230, and CB~244. These six globule cores are the strongest millimeter continuum sources
from surveys by LH97 and Henning \& Launhardt~(1998) surveys accessible with the JCMT/SCUBA.
They are the best-studied star-forming
Bok-globules.


B\,335 (CB\,199), an isolated, nearly spherical Bok globule at a distance of 
$\sim$\,250\,pc (Tomita, Saito, \& Ohtani~1979; Frerking, Langer, \& Wilson~1987). 
It accommodates one of the best-studied low-mass protostellar cores 
(see, e.g., Myers et al.~2000). The deeply embedded Class\,0 protostar of $L_{\rm bol} \sim
3$\,L$_{\odot}$\ drives a collimated bipolar outflow with a dynamical age of $\sim
3\times 10^4$\,yr  (Keene et al.~1983; Hirano et al.~1988; Cabrit, Goldsmith, \& Snell~1988; 
Chandler et al.~1990, Chandler \& Sargent~1993).
The dense core in B\,335 is generally recognized as the best protostellar 
collapse candidate and the emission from different molecular lines has
been  successfully modeled in terms of an inside-out collapse (Shu 1977) with
infall age $\sim 10^5$\,yr and a current protostar mass of $\sim 0.4$\,M$_{\odot}$\ 
(Zhou et al.~1993; Choi et al.~1995).
Recent molecular line observations, made with higher angular resolution,
show possible discrepancies with the predictions of
inside-out collapse (Wilner et al.~2000). From our SCUBA maps, we derive a total envelope mass 
of $\sim$\,4\,M$_{\odot}$\ within a radius of $1.5\times 10^4$\,AU
(Launhardt et al.~in prep.). 
The small C$^{18}$O line widths, observed by Frerking et al.~(1987), imply a 
turbulent velocity dispersion of only $0.14\pm0.02$\,km\,s$^{-1}$\ in the dense core. 

CB230 (L\,1177) is a small, bright-rimmed Bok globule associated with
the Cepheus Flare molecular cloud complex. While in earlier papers we
suggested a distance of 450\,pc (LH97), Kun~(1998) pointed
out that L\,1177 is more likely associated with an absorbing sheet at 300\,pc. 
We therefore use a distance of 400$\pm$100\,pc. 
The globule contains a binary protostellar core with 10\arcsec\ separation (east-west) 
with signatures of mass infall which is associated with a double NIR reflection nebula. 
Both protostars are associated with a NIR reflection nebula and they drive separate 
aligned molecular outflows, but only the western one is associated with a 
massive accretion disk (Launhardt et al.~1998; Launhardt~2001). 
The total mass and luminosity of the protostellar double core, which is
unresolved in our SCUBA maps, are $\sim$\,7\,M$_{\odot}$\ within a radius of
$2\times 10^4$\,AU and $\sim$\,10\,L$_{\odot}$, respectively 
(Launhardt~2001; Launhardt et al.~in prep.). The dynamical age of the collimated 
large-scale outflow is $\sim 2\times 10^4$\,yr (Yun \& Clemens~1994).
From the observed C$^{18}$O line width of 0.7$\pm$0.1\,km\,s$^{-1}$\ 
(Wang et al.~1995; Launhardt~1996) we calculate an upper limit for the turbulent 
velocity dispersion of $0.29\pm0.04$\,km\,s$^{-1}$. 

CB\,244 (L\,1262) is an isolated Bok globule located at a distance of 180\,pc and 
probably associated with the Lindblad ring (LH97; Kun~1998). 
It contains two dense cores separated by $\sim$\,\,90\arcsec. The more prominent 
south-eastern core, which we observed here, contains a Class\,0 protostar with 
signatures of mass infall. It is associated with a bipolar molecular outflow 
with a dynamical age of $\sim 10^4$\,yr (Yun \& Clemens~1994; Wang et al.~1995; 
Launhardt~1996; LH97; Launhardt et al.~1997).
The total mass and bolometric luminosity of this protostellar core are 
$\sim$\,2\,M$_{\odot}$\ within a radius of $1.8\times 10^{4}$\,AU and $\sim$\,1.5\,L$_{\odot}$, 
respectively (Wang et al.~1995; Launhardt et al.~in prep.). 
The C$^{18}$O line width and turbulent velocity dispersion are the same as for 
CB\,230, i.e., $\sigma_{\rm turb}\sim 0.29$\,km\,s$^{-1}$.

\begin{deluxetable}{llcccc}
\tablecaption{Coordinates and distances of the observed globules.\label{coordinates}}
\tablehead{
\colhead{Source name\tablenotemark{(a)}}    & 
\colhead{Other}                             & 
\colhead{IRAS}                              & 
\colhead{R.A.~(B1950)}                      & 
\colhead{Dec.~(B1950)}                      &  
\colhead{Dist.}                             \\
\colhead{}                                  &
\colhead{names\tablenotemark{(a)}}          &
\colhead{source}                            & 
\colhead{[\ h\ m\ s\ ]}                     & 
\colhead{[\ \degr\ \ \arcmin\ \ \arcsec\ ]} & 
\colhead{[pc]}                              }
\startdata
B\,335            &  CB\,199  &       --       &  19 34 35.1$^{\rm b}$  &  + 07 27 20$^{\rm b}$  &  250$^{\rm d}$ \\
CB\,230           &  L\,1177  &  21169$+$6804  &  21 16 53.7$^{\rm c}$  &  + 68 04 55$^{\rm c}$  &  400 \\
CB\,244 (SE core) &  L\,1262  &  23238$+$7401  &  23 23 48.5$^{\rm c}$  &  + 74 01 08$^{\rm c}$  &  180$^{\rm e}$ \\ 
\enddata
\tablenotetext{(a)} {B: Barnard 1927; L: Lynds 1962; CB: Clemens \& Barvainis~1988.}
\tablenotetext{(b)} {Coordinates of the submillimeter peak (Huard et al.\ 1999).}
\tablenotetext{(c)} {Coordinates of the 1.3~mm peak (Launhardt et al.~in prep.).}
\tablenotetext{(d)} {Tomita et al.\ 1979.}
\tablenotetext{(e)} {LH97 (for CB~230 see source description).}
\end{deluxetable}

\begin{deluxetable}{lcccc}
\tablecaption{Total and peak fluxes of the observed globules (from Launhardt et al.~in prep.).\label{fluxes}}
\tablehead{
\colhead{Object}                  & 
\colhead{$F_{\rm 450}$}           &
\colhead{$F_{\rm 850}$}           &
\colhead{$I^{\rm peak}_{\rm 450}$} &
\colhead{$I^{\rm peak}_{\rm 850}$} \\
\colhead{}                        &
\colhead{[Jy]}                    & 
\colhead{[Jy]}                    &
\colhead{[Jy/$\Omega_{\rm b}$]\tablenotemark{(a)}} & 
\colhead{[Jy/$\Omega_{\rm b}$]\tablenotemark{(a)}} }
\startdata
B\,335 (CB\,199)  &  --    &  3.7 & --   & 1.15 \\
CB\,230           &  16.55 &  2.9 & 3.86 & 0.92 \\
CB\,244 (SE core) &  9.1   &  1.8 & 1.69 & 0.49 \\ 
\enddata
\tablenotetext{(a)} {Effective beam sizes (HPBW) $\Omega_{\rm b}$ derived from calibrator maps are 
 8.5'' at 450\,$\mu$m and
14.4'' at 850\,$\mu$m.}
\end{deluxetable}

\section{Observations and Data reduction} \label{obs}

The observations were performed at the 15-m JCMT on Mauna Kea (Hawaii)
between September 10 and 14, 2001.
The effective beam size (HPBW) at 850$\mu$m is $\sim$\,14.4\arcsec\ at 850\,$\mu$m.
Polarimetry was conducted using SCUBA (Holland et al.~1999) and its polarimeter, 
SCU-POL with the 350-850 $\mu$m achromatic half-waveplate.
For a detailed description of the polarimeter hardware
we refer to Murray et al.\ (1997) and Greaves et al.\ (2000).

Since our targets have a protostellar core/disk-envelope structure with 
envelope sizes smaller than 2\arcmin, we used the imaging mode of SCUBA.
Fully sampled 16-point jiggle maps have been obtained for each object, whereby each jiggle map
was repeated 16 times with the wave plate turned by 22.5\degr\ between the individual maps.
This mode allows simultaneous imaging polarimetry with a 2.3~arcminute field of view
in the long (750/850~$\mu$m) and short (350/450~$\mu$m) wavelength bands.
However, only 850~$\mu$m data are presented in this paper because the signal-to-noise ratio
was too low for the 450~$\mu$m polarimetry data.

The data reduction package SURF (SCUBA User Reduction Facility; see Jenness \& Lightfoot~1998)
was used for flat-fielding, extinction correction, sky-noise removal 
(see Jenness et al.~1998), and instrumental polarization correction. 
The Stokes parameters I, Q, and U were computed for each set of the 16 maps  using the POLPACK
data reduction package (Berry \& Gledhill 1999) by averaging maps taken
at the same wave plate orientation followed by fitting a sine wave to each image pixel.
This set of Stokes parameters was then averaged and binned (over a 9\arcsec\ region)
before calculating the average linear polarization degree $P_{\rm l}$ and position angle
$\gamma$ for each pixel.

Since the chop throw was 120", the very outer regions 
in the jiggle maps (which are also undersampled) may suffer 
from chopping into the outermost envelope regions 
and into extended low-level emission from the tenuous outer regions of the globules.
We therefore restrict the polarization analysis to the 
inner region with a radius of 60\arcsec\ of the maps and do not use the outer $\sim$20\arcsec.
Based on the 1.2\,mm continuum maps obtained by Launhardt et al.~(in prep.),
we chose arbitrarily a chop throw (azimuthal direction) for
B~335 and CB~230, while CB~244 was observed chopping almost perpendicular to 
the axis defined by the main and the secondary component, in order to avoid the secondary component
(see Fig.~\ref{allpolpat} for illustration).

We restrict the polarization analysis to regions in which the 
total flux density is higher than 5 times the rms in the maps 
(measured outside the central sources).
Pixels in which the scatter of the total flux density measurements between 
different jiggle cycles was larger than 20\% of the average value were also excluded.
Furthermore, polarization vectors with $P_{\rm l}/\sigma(P_{\rm l}) >$ 3,
where $\sigma(P_{\rm l})$ is the standard deviation of the polarization degree, have been excluded.
We found that our selection criteria, which are based on
signal-to-noise
ratios, result in polarization maps that agree well with those
obtained with the default restrictions of ORACDR/POLPACK which
use absolute thresholds ($0.1\% < P_{\rm l} < 15\%$ and $\sigma_{\gamma}<10^{\rm o}$).

\section{Results}\label{results}

\subsection{Polarization maps} \label{polpat}

\begin{figure}
\begin{center}
  \resizebox{6.5cm}{!}{\includegraphics{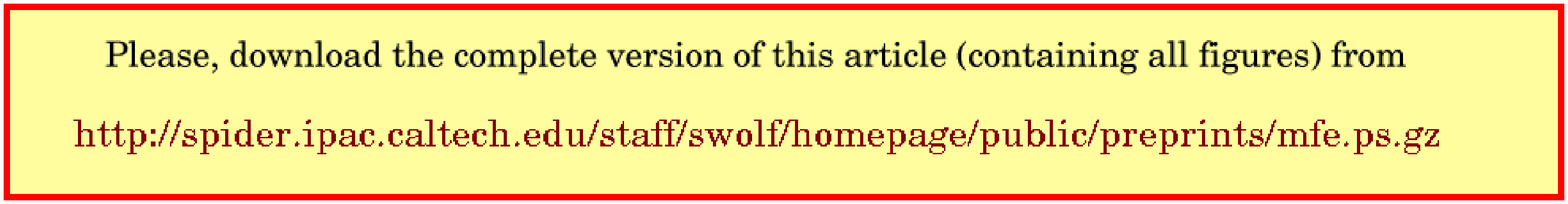}}\\
  \resizebox{6.5cm}{!}{\includegraphics{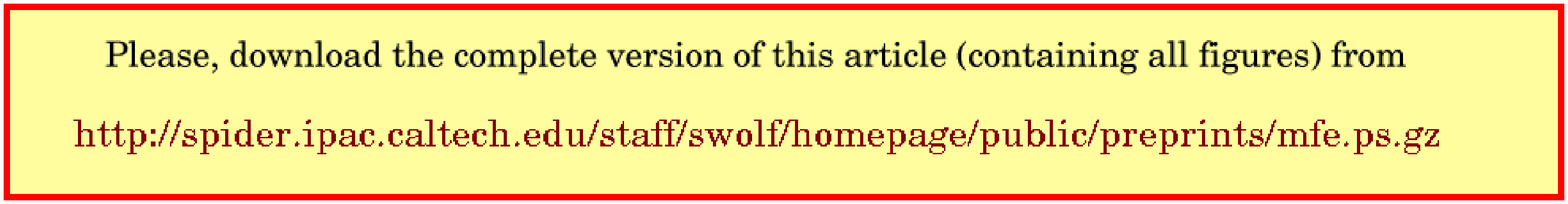}}\\
  \resizebox{6.5cm}{!}{\includegraphics{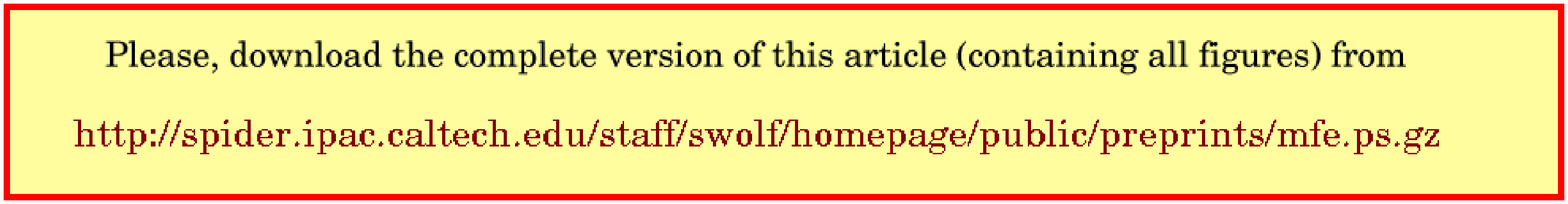}}
\end{center}
  \caption[]{850~$\mu$m SCUBA maps of B~335 (top), CB~230 (middle), CB~244 (bottom)
	and with polarization vectors superimposed.
	The corresponding right ascension and declination of the (0,0) coordinate
	are given in Tab.~\ref{coordinates}.
	The length of the vectors is proportional to the degree of polarization,
	and the direction gives the position angle. The data are binned over 9''.
	Only vectors for which the 850~$\mu$m flux exceeds 5 times the standard
	deviation and $P_{\rm l}/\sigma(P_{\rm l})>3$ are plotted.
	The contour lines mark the levels of 20\%, 40\%, 60\%, and 80\%
	of the maximum intensity.
	In case of CB~244, the time-dependent chopping direction during the observation 
	is symbolized in the upper left edge.
    }
  \label{allpolpat}
\end{figure}

Polarized thermal emission by aligned non-spherical
grains is the main source of polarized submillimeter radiation 
in Bok globules (see, e.g., Weintraub et al.~2000, Greaves et al.~1999).
The polarization maps of the Bok globules B~335, CB~230, and CB~244 at 850~$\mu$m
are shown in Fig.~\ref{allpolpat}.
\begin{figure*} 
  \resizebox{\hsize}{!}{\includegraphics{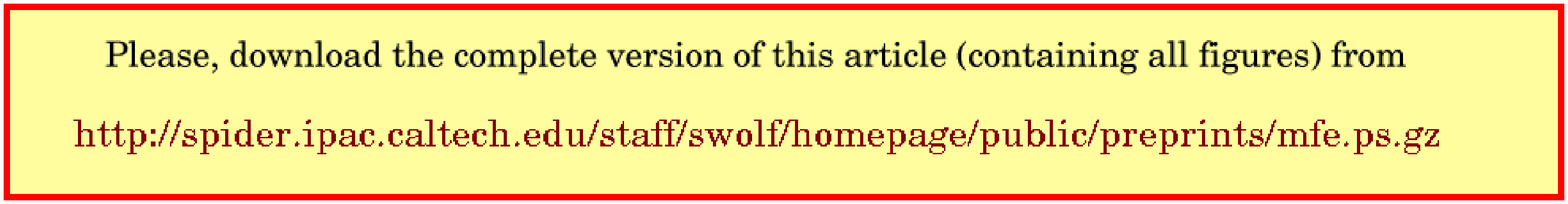}}
  \caption[]{Histograms showing the distribution of the degree of polarization
    around the Bok globules B~335, CB~230, and CB~244.
    }
  \label{plhis}
\end{figure*}
In Figure~\ref{plhis} we plot the polarization histogram for each globule.
The degree of linear polarization reaches values up to 14\%, whereby the distributions $N(P_{\rm l})$ 
is strongly influenced by statistical noise due to the small number of data points.
The mean percentage polarization degrees for B~335, CB~230, and CB~240 are 
6.2\%, 
7.8\%, 
and 
5.1\%. 
The corresponding 1$\sigma$ dispersions are 
3.7\%, 
4.1\%,
and 
3.6\%.
These values are similar to those published in Paper\,I
for the Bok globules CB~26, CB~54, and CG~30 
(for polarization measurements in larger molecular clouds
see
Dowell~1997, 
Novak et al.~1997,
Hildebrand~1996,
Hildebrand et al.~1990, 1993,
Morris et al.~1992, and
Gonatas et al.~1990).

\subsection{Magnetic fields} \label{magfield}

An estimate of the magnetic field strength (in units of G) can be derived from
the polarization maps as follows (see Chandrasekhar \& Fermi 1953)
\begin{equation}\label{eqb}
B = |\vec{B}| = \sqrt{\frac{4\,\pi}{3} \, \rho_{\rm Gas}} 
\cdot \frac{v_{\rm turb}}{\sigma_{\bar \gamma}}\ .
\end{equation}
Here, $\rho_{\rm Gas}$ is the gas density (in units of g\,$\rm cm^{-3}$),
$v_{\rm turb}$ the rms turbulence velocity (in units of cm\,$\rm s^{-1}$), and
$\sigma_{\bar{\gamma}}$ the standard deviation to the mean orientation 
angle $\bar{\gamma}$ of the polarization vectors (in units of radians).
Hereby, it is assumed that the magnetic field is frozen in the cloud material.
For a detailed discussion of the applicability of this equation we refer
to Paper\,I (\S5.3).

\begin{deluxetable}{lcccccccccc}
\tablecaption{Masses, gas densities, polarization, and magnetic field
strengths of the envelopes.\label{obres}}
\tablehead{
\colhead{Object}                    &
\colhead{$M_{\rm H}^{\rm env}$}     & 
\colhead{$\langle n_{\rm H}\rangle$}& 
\colhead{$\rho_{\rm Gas}$}          & 
\colhead{$v_{\rm turb}$}            & 
\colhead{$N_{\rm vec}$}             & 
\colhead{$\bar{\gamma}$}            & 
\colhead{$\sigma_{\bar{\gamma}}$}   & 
\colhead{B}                         \\
\colhead{}                          & 
\colhead{[M$_{\odot}$]}             &
\colhead{[${\rm cm^{-3}}$]}         & 
\colhead{[g\,${\rm cm^{-3}}]$}      & 
\colhead{[km\,${\rm s^{-1}}]$}      &
\colhead{}                          & 
\colhead{[\ \degr \ ]}              & 
\colhead{[\ \degr\ ]}               & 
\colhead{[$\mu$G]}                
}
\startdata
B\,335                   & 5    & 3.8E+6 & 8.6E-18 &  0.14$\pm$0.02$^{\rm a}$ &  20  &  -87.0  &  $35.8^{+14.6}_{-9.1}$   &  134$^{+46}_{-39}$  \\
CB\,230                  & 7    & 1.6E+6 & 3.6E-18 &  0.29$\pm$0.04$^{\rm b}$ &  33  &   23.5  &  $29.8^{+8.8}_{-6.1}$    &  218$^{+56}_{-50}$  \\
CB\,244 (SE core)        & 1.5  & 3.5E+6 & 8.0E-18 &  $\approx$0.29$^{\rm c}$ &  12  &   68.3  &  $33.1^{+19.2}_{-10.4}$  &  257$^{+111}_{-91}$ \\
CB\,26$^{\rm e}$         & 0.27 & 3.8E+5 & 8.6E-19 &           0.25$^{\rm d}$ &   7  &   25.3  &  $18.9^{+16.7}_{-7.3}$   &  144$^{+91}_{-68}$  \\
CB\,54$^{\rm e}$         & 100  & 1.5E+5 & 3.4E-19 &           0.65$^{\rm c}$ &  41  &  -68.0  &  $42.7^{+11.1}_{-8.0}$   &  104$^{+24}_{-21}$  \\
DC\,253-1.6$^{\rm e,f}$  & 9    & 2.2E+6 & 5.0E-18 &           0.25$^{\rm d}$ &  49  &   14.4  &  $38.2^{+8.9}_{-6.6}$    &  172$^{+36}_{-33}$  \\
\enddata
\tablenotetext{(a)}{Frerking et al.~1987.}
\tablenotetext{(b)}{Wang et al.~1995; Launhardt et al.~1996.}
\tablenotetext{(c)}{Wang et al.~1995}
\tablenotetext{(d)}{No direct value available. rms turbulence velocity of a large sample of nearby star-forming Bok globules derived from C$^{18}$O (J=2-1).} 
\tablenotetext{(e)}{Re-derived mean density $\langle n_{\rm H}\rangle$ and corrected value for the magnetic field strength B.}
\tablenotetext{(f)}{The magnetic field given for DC~253-1.6 in Tab.~2 of Paper\,I has to be corrected to $58^{+12}_{-11}\mu$G.}
\end{deluxetable}

Since the gas density $\rho_{\rm Gas}$ obviously strongly increases towards
the center of each core, the magnetic field strength should be derived as
a function of the radial distance from the density center. 
However, the small number of polarization vectors fulfilling the selection criteria
discussed in \S\ref{obs} (see also Tab.~\ref{obres}) 
does not allow us to perform a statistical analysis
on individual subsamples of polarization data points.
Therefore, we decided to base our magnetic field estimates on the mean
density from which the 850\,$\mu$m emission at a certain
projected distance from the emission center arises.
To ensure that the density value represents the region from
which $\bar{\sigma}_\gamma$ was calculated, we used the mean distance of
all considered polarization vectors from the center.
These angular distances are
B~335:  16.6'',
CB~230: 21.9'', and
CB~244 (SE core): 13.1''.

We calculate the hydrogen number density profiles by using a
ray-tracing code to fit spherically symmetric source models with
an outer power-law density gradient to the observed, circularly
averaged intensity maps. The model maps were convolved with the
observed beam shape and chopping was accounted for.
The mean dust temperature in the envelopes was determined by the
450/850\,$\mu$m surface brightness ratio under
the assumption that the dust opacity index
$\beta = -1.8$\ and
$\kappa_{\rm dust}$(1.3\,mm)\,=\,0.5\,cm$^2$\,g$^{-1}$.
Dust temperatures in the range 10 to 14\,K with an unresolved
warmer core were derived, except for CB\,26 and CB\,54, which
have higher temperatures and a global radial temperature gradient.
Details of the models will be given in a forthcoming paper
(Launhardt et al.~in prep.).
The derived envelope dust temperatures agree well with the kinetic
gas temperatures derived by Cecchi-Pestellini et al.~(2001) for a
number of very similar southern Bok globules with protostellar cores.
They also agree with the outer envelope temperature Evans et al.~(2001)
calculate for star-less globules which are heated interstellar radiation
field only.

The dust emissivity is converted into hydrogen number density by using a standard
hydrogen-to-dust mass ratio of 100. Details of the model will be
given in a forthcoming paper (Launhardt et al.~in prep.).
To account for helium and heavy elements, we derive the total gas density $\rho_{\rm Gas}$ by
\begin{equation}\label{rhon}
  \rho_{\rm Gas} = 1.36 \cdot n_{\rm H} \cdot M_{\rm H},
\end{equation}
where $M_{\rm H} = 1.00797$\,mu is the mass of a hydrogen atom.
The corresponding densities are listed in Tab.~\ref{obres}.

\begin{figure*} 
  \resizebox{\hsize}{!}{\includegraphics{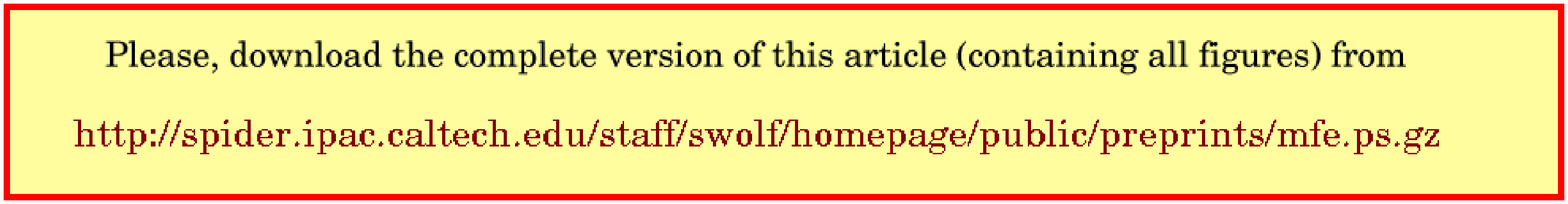}}
  \caption[]{Histograms showing the distribution of position angles
    around B~335, CB~230, and CB~244.
    Only data points in which the 850~$\mu$m flux exceeds 5 times
    the standard deviation and $P_{\rm l}/\sigma(P_{\rm l}) > 3$ have been considered.}
  \label{orihis}
\end{figure*}
\begin{figure} 
  \resizebox{0.8\hsize}{!}{\includegraphics{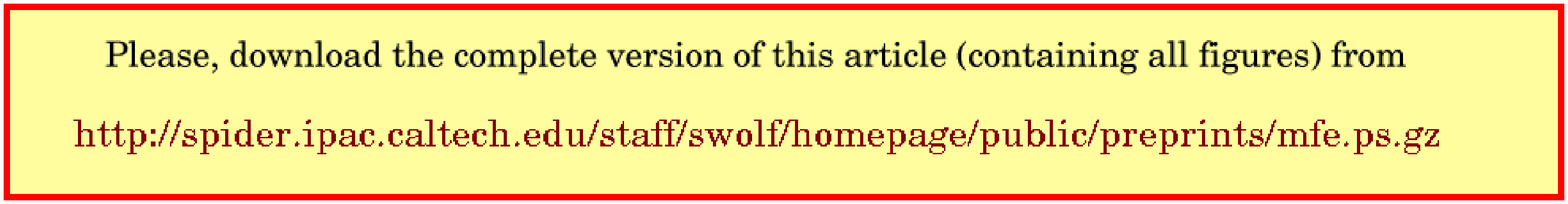}}
  \caption[]{
	{\em Top:}
	Mean orientation angle $\bar{\gamma}$ of the polarization vectors
    	as a function of the number of combined maps in the case of CB~230.
  	The error bars mark the range $\bar{\gamma} \pm \sigma_{\bar{\gamma}}$.
	{\em Bottom:}
	Standard deviation to the orientation angle of the net polarization
	as a function of the number of combined maps in the case of CB~230.}
  \label{conv-sig}
\end{figure}

The next quantity to be derived from the polarization map in order
to achieve an estimate of the magnetic field strength is the
standard deviation of the orientation angle $\sigma_{\bar{\gamma}}$.
The histograms of the orientation angle and the resulting standard deviations
are shown in Fig.~\ref{orihis} and compiled in Tab.~\ref{obres}.
Using CB~230 as an example, Fig.~\ref{conv-sig} also illustrates that a clear convergence
towards a constant value of both quantities could be achieved
on the basis of 18 complete polarization measurement cycles (``exposures'').
For comparison, 27/15 exposures have been obtained for the globules B~335
and CB~244, respectively.

The error estimates for the value of the standard deviation of the mean orientation angle
$\sigma_{\bar{\gamma}}$ is based on a $\chi^2$ test assuming a standard Gaussian 
distribution of the orientations of the polarization vectors. 
The error intervals given in Tab.~\ref{obres} are based on confidence intervals for
$\sigma_{\bar{\gamma}}$.
The probability for the real (unknown) $\sigma_{\bar{\gamma}{\rm, real}}$ 
to be included in this interval amounts to 95\%.
Based on this error estimate for $\sigma_{\bar{\gamma}}$, we give error intervals
for the magnetic field strength $B$.

The magnetic fields -- determined by the application of Equation~\ref{eqb} -- are 
$B_{\rm B335}  \approx 130\,\mu$G,
$B_{\rm CB230} \approx 220\,\mu$G, and
$B_{\rm CB244} \approx 260\,\mu$G.
According to the investigations of theoretical models of polarized dust emission
from protostellar cores by Padoan et al.~(2001), these values might have to be corrected by a factor
of $f \approx 0.4$ in order to provide a better estimate of the average
magnetic field strength in the cores.

In Paper\,I, we used volume-averaged densities, which are by a factor of about 3 (CB\,54), 4 (CB\,26), 
and 9 (DC\,253-1.6) lower than those derived with our method applied here.
For the sake of consistency and comparability, we re-derive densities
and magnetic field strengths for these three sources and give the corrected values in Tab.~\ref{obres}.
The magnetic field strengths in all six considered Bok globules
are therefore very similar, amounting to $\approx$ 0.1-0.3 \,mG.
These magnetic field strengths are in the range of those values found in 
molecular clouds, pre-protostellar cores, and other star-forming regions
(see, e.g.,
Matthews \& Wilson~2002, 
Levin et al.~2001,
Davis et al.~2000,
Crutcher~1999,
Glenn et al.~1999,
Itoh et al.~1999,
Minchin \& Murray~1994,
Chrysostomou et al.~1994,
Bhatt \& Jain~1992).

\subsection{$P_{\rm l}$ vs. $I$ behavior}\label{plvsi}

\begin{figure*}
  \resizebox{0.6\hsize}{!}{\includegraphics{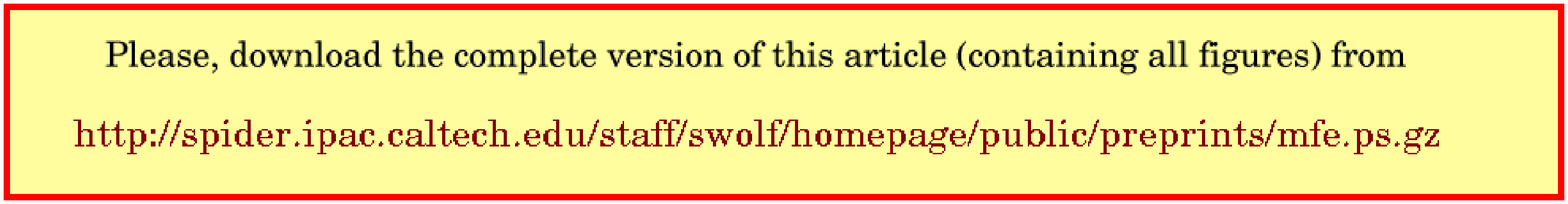}}
  \caption[]{Scatter diagrams showing the distribution of $P_{\rm l}$ vs.\ intensity $I$ across
    the Bok globules B~335, CB~230, and CB~240. Only data in which the 850~$\mu$m flux exceeds 5 times
    the standard deviation, and $P_{\rm l}/\sigma(P_{\rm l}) > 3$ have been considered. 
    Fits to the data sets of the data described by functions of the form
    $P_{\rm l} = a_0 + a_1 (I/{\rm max}(I))^{a_2}$ are superposed
    on the data points (see \S\ref{plvsi}).
    }
  \label{scatt}
\end{figure*}
Similarly to previous polarization measurements in other star-forming cores
(see, e.g., Matthews \& Wilson~2002; Houde et al.~2002; 
Minchin et al.~1996; Glenn et al.~1999; see also Paper\,I),
the degree of polarization was found to decrease towards regions of increasing intensity 
(see Fig.~\ref{scatt}).
This behaviour can be explained by either
($i$) an increase of the density in the brighter cores,
resulting in an increased collisional disalignment rate of the grains towards the centers 
of the cores, 
($ii$) grain growth in the denser regions resulting in unpolarized 
re-emission by the dust (Weintraub et al.~2000), or 
($iii$) the fact that
the field structure associated with the core collapse may be still unresolved 
in our polarization maps (see, e.g., Shu et al.~1987).

As outlined in Paper\,I, the decrease of the polarization
towards increasing intensity can be approximately described by
\begin{equation}\label{nonlinfit}
  P_{\rm l} = a_0 + a_1 \left(\frac{I}{{\rm max}(I)}\right)^{a_2},
\end{equation}
where $P_{\rm l}$ is the degree of linear polarization, $I$ is the measured intensity,
and $a_0$, $a_1$, and $a_2$ are constant quantities.

\begin{figure}
  \resizebox{\hsize}{!}{\includegraphics{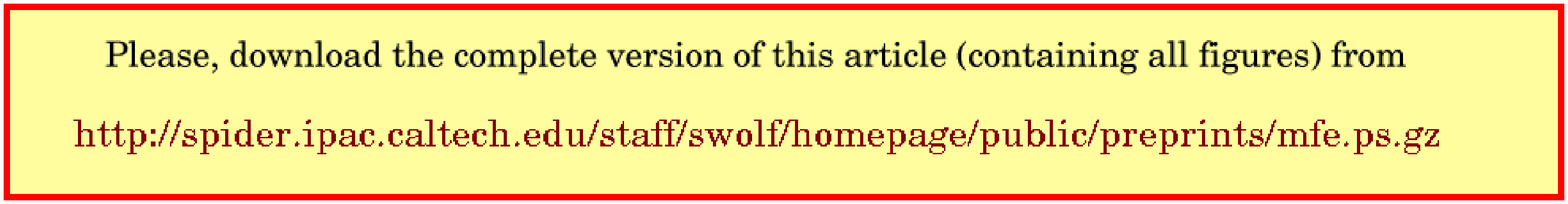}}
  \caption[]{
    {\em Left:}
    Scatter diagram showing the distribution of $P_{\rm l}$ 
    vs.\ intensity $I$ across the Bok globules 
    B~335, 
    CG~30, 
    CB~54, 
    CB~230,
    CB~240, and
    CB~244 (SE core). 
    The data for the Bok globules CG~30, CB~26, and CB~54
    have been taken from Paper\,I.
    The solid line represents the best fit of the function
    given in Eq.~\ref{nonlinfit}.
    {\em Right:}
    Corresponding distribution of the polarization degree
    as a function of the hydrogen density (see \S\ref{plvsi}).
    If the measurements for CB~54 (symbolized by triangles) were excluded from this plot,
    a nearly linear dependency between the polarization degree and gas density
    is reaveled for densities $\log_{10}n_{\rm H} < 8$. What distinguishes CB~54 from the other objects
    is the fact that it most likely contains several unresolved sources (see Paper~I for a detailed
    description of this source).
   }
  \label{allfits}
\end{figure}
Under the assumption that the physical properties of the dust grains are similar
in the three Bok globules considered in this paper as well as in the Bok globules
CG~30, CB~26, and CB~54 investigated in Paper\,I, the data points
(162 in total) can be combined in order to derive a significantly better
functional relationship between the intensity and polarization (and therefore
the magnetic field strength than we could give in Paper~I). 
To combine the data, the six separate intensity distributions have been
normalized according to Eq.~\ref{nonlinfit}.
We find the following parameters:
$a_0$ = -1.70, $a_1$ = 3.96, $a_2$ = -0.43
for the best fit of Eq.~\ref{nonlinfit} to the total data set (see Fig.~\ref{allfits}).
In fact, the average fit values for the entire sample are very similar to those found for CB~54 and DC~253-1.6 (see Paper~I).

We remark that Eq.~\ref{nonlinfit} represents an ad-hoc assumption
about the relation between the polarization degree $P_{\rm l}$ and the corresponding 
intensity $I$, introduced to allow a first-order quantitative comparison
of this relation for different Bok globules. 
A similar dependency $P_{\rm l}(I)$ was also found for other Bok globules
and star-forming regions (Matthews \& Wilson~2002; Houde et al.~2002; Minchin et al.~1996;
Glenn et al.~1999; see also Paper\,I).
A first qualitative confirmation of this observationally found non-linear dependency
was provided by Padoan et al.~(2001) on the basis of MHD simulations assuming no alignment of grains
in regions with $A_{\rm V} > 3$\,mag.
A dependency in the form $P( n(\vec{r}), v_{\rm turb}(\vec{r}))$ is in fact expected if
the decrease of the polarization degree is due to an increased disalignment rate
of the dust grains in regions of high density and turbulence velocity.

\subsection{Correlation between the magnetic field structure and the morphology of the Bok globules}\label{magmorph}

All three globules (B~335, CB~230, and CB~244) contain Class 0 protostellar cores and drive
collimated bipolar molecular outflows (Chandler \& Sargent~1993, Yun \& Clemens~1994, Launhardt~2001).
Our polarization maps reveal alignment of polarization vectors and therefore the magnetic field 
with the outflow direction (most prominent in the case of B~335).
Thus, the question arises whether the outflow direction is somehow related to the magnetic
field structure given by the polarization maps. 
Since for two of our previously investigated globules
(CB~26 and CB~54, see Paper\,I) we also know the outflow direction, we include these sources in our discussion
here. The double core in DC\,253-1.6 drives two nearly perpendicular outflows and is, therefore, not considered here.
In discussing the relative orientation and relation between the magnetic field and the outflow
one has to consider that only one component of the spatial orientation of the outflow ($v \sin i$) 
is known from velocity measurements and the polarization
vectors only allow to trace the projection of the magnetic field on the plane of the sky.
In our analysis we assume that the magnetic field is oriented perpendicular
to the measured polarization pattern. This widely applied concept is based on the finding
that irrespective of the alignment mechanism, charged interstellar grains would have a
substantial magnetic moment, leading to a rapid precession of the grain angular
momentum $\vec{J}$ around the magnetic field direction $\vec{B}$ which implies a net alignment 
of the grains with the magnetic field (see, e.g., Draine \& Weingartner 1997).

The 850\,$\mu$m polarization maps overlayed with the blue- and redshifted outflow velocity
contour lines of B~335, CB~230, CB~244, CB~26, and CB~54 are shown in Fig.~\ref{of-b335}-\ref{of-cb26}.
Since the outflows have been mapped using different radio telescopes, the spatial resolution
varies strongly but the main features, i.e., the orientation with respect to 
the aligned polarization vectors can be clearly seen:

\begin{figure*}
  \resizebox{\hsize}{!}{\includegraphics{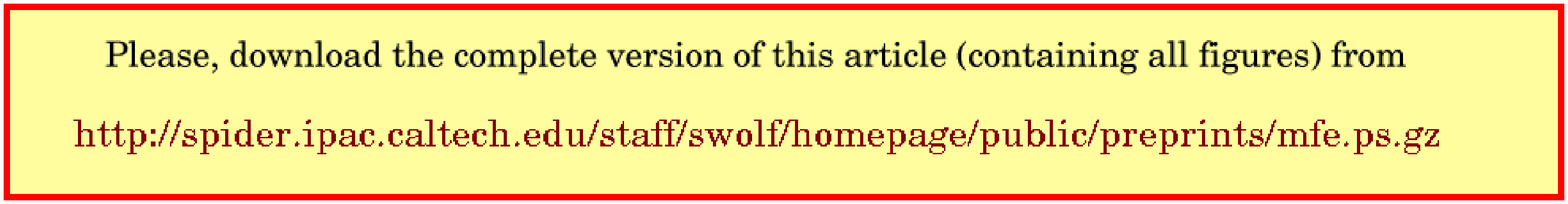}}
  \caption[]{SCUBA intensity map (850\,$\mu$m) of the Bok globules B~335 and CB~230 with overlayed
    \begin{enumerate}
    \item {\em Iso-intensity contour lines} in the case of CB~230 
      (thin dotted lines; 850\,$\mu$m; contour levels: 
      0.15\%, 0.30\%, 0.45\%, 0.60\%, 0.75\%, and 0.90\% of the maximum intensity $I_{\rm max}$.
      For reason of clarity, iso-intensity contour lines corresponding
      to the 850\,$\mu$m SCUBA intensity map are not shown in the case of B~335 -- 
      see Fig.~\ref{allpolpat}(top) for this information.),
    \item {\em Polarization pattern} (850\,$\mu$m, see Fig.~\ref{allpolpat} for details),
	and
    \item {\em $^{13}$CO(1-0) spectral channel maps} obtained with OVRO.
      {\bf B~335}:
      The white/grey contour lines represent the spatially well-separated
      blue/red shifted western/eastern outflow lobe. The contours are spaced at $2\sigma$ intervals
      of 200\,mJy\,beam$^{-1}$ (from Chandler \& Sargent~1993; beam width: 2.9''$\times$2.9'').
      {\bf CB~230}:
      The solid/dashed contour lines represent the spatially well-separated
      blue-shifted/red-shifted outflow lobe (beam width: 4'').
      Blue lobe: $v_{\rm LSR}=1.6\ldots$2.8\,km\,s$^{-1}$ (solid contours),
      red  lobe: $v_{\rm LSR}=3.0\ldots$4.2\,km\,s$^{-1}$ (dashed contours).
      The step width amounts to 0.3\,km\,s$^{-1}$ for both lobes (from Launhardt~2001).
      See also Yun \& Clemens~(1994, Fig.~27) for a large-scale outflow map ($6'\times6'$).
    \end{enumerate}
    }
  \label{of-b335}
\end{figure*}

\noindent{\bf B~335} (Fig.~\ref{of-b335}, left):
Nearly {\em parallel} alignment of the polarization vectors 
(and therefore the mean polarization) with the outflow axis.
The globule is clearly elongated in North-South direction (see Fig.~\ref{allpolpat}), and therefore
oriented perpendicular to the polarization pattern and parallel to the magnetic field.
Based on investigations of axis ratios of large samples of Bok globules,
Myers et al.~(1991) and Ryden~(1996) found that oblate cores 
would be inconsistent with the observed axis ratios to a high confidence level.
Thus, we assume that B~335 is a prolate globule rather than an oblate one seen edge-on.
The orientation of the magnetic field parallel to the symmetry axis of the globule
fits into the scenario described by Fiege \& Pudritz~(2000a; 
see also Fiege \& Pudritz~2000c), assuming the case 
$B_{\rm z}/B_{\rm \Phi} > 0.37$, where 
$B_{\rm z}$ and
$B_{\rm \Phi}$ are, respectively, 
the ploidal and toroidal magnetic field components at the outer surface of the globule core.
Their theoretical investigations of molecular cloud cores that originate from filamentary clouds which
are threaded by helical magnetic fields show that the radial pinch of the toroidal field
component helps to squeeze cores radially into a prolate shape while helping to support
the gas along the axis of symmetry. In the context of star formation inside Bok globules
it is of interest that the Bonnort-Ebert critical mass is reduced by about 20\% 
by the toroidal field. 
Assuming that the predicted submillimeter polarization patterns for filamentary clouds
are valid for elongated Bok globules as well, we can not completely confirm the predictions
by Fiege \& Pudritz~2000b (see Fig.~1 in their work for different 
magnetic field/polarization pattern scenarios).
Based on model simulations, these authors find
depolarization along the axis of the filaments. While we find depolarization towards
the center of the Bok globules as well (see \S\ref{plvsi} for a detailed analysis of this
effect), the radial dependence of the polarization degree in the outer regions of
the Bok globules does not agree with the theoretical findings for filamentary structures.

\noindent{\bf CB~230} (Fig.~\ref{of-b335}, right):
Alignment of the mean polarization almost {\em parallel} to the outflow axis.
The decrease of the linear polarization towards the bright globule core 
is slightly stronger towards the red-shifted side of the lobe. 
The globule is slightly elongated in E-W direction and therefore, like in the case
of B~335 (under the assumption of a spheroidal shape) elongated perpendicular 
to the polarization pattern and parallel to the magnetic field direction, respectively.

\begin{figure*}
  \resizebox{\hsize}{!}{\includegraphics{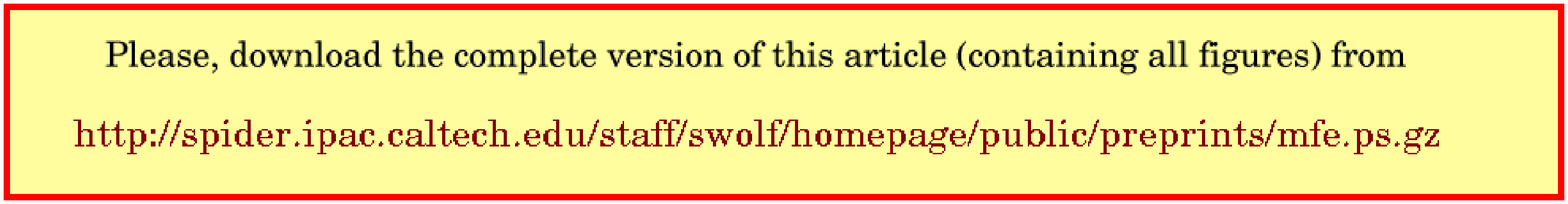}}
  \caption[]{SCUBA intensity map (850\,$\mu$m) of the Bok globule CB~244 and CB~54 with overlayed
    \begin{enumerate}
    \item {\em Iso-intensity contour lines} (thin solid lines; 850\,$\mu$m; contour levels: 
      0.15\%, 0.30\%, 0.45\%, 0.60\%, 0.75\%, and 0.90\% of the maximum intensity $I_{\rm max}$),
    \item {\em Polarization pattern} (850\,$\mu$m, see Fig.~\ref{allpolpat} for details), and
    \item {\em $^{12}$CO $J$=1-0 spectral line map} (from Yun \& Clemens~1994).
      The broad solid/dashed contour lines represent the blue/red integrated
      line wing emission.
      These maps were obtained with a 15 beam receiver with an antenna had beam width of 48'' (FWHM;
      for comparison: SCUBA FWHM amounts to 14.7'' at 850\,$\mu$m).
      {\bf CB~244}:
      Blue/red-shifted outflow contour lines  begin with 0.5\,K\,km\,s$^{-1}$/0.6\,K\,km\,s$^{-1}$
      and are stepped by 0.15\,K\,km\,s$^{-1}$/0.2\,K\,km\,s$^{-1}$.
      {\bf CB~54}:
      Blue/red-shifted outflow contour lines begin with 1.7\,K\,km\,s$^{-1}$/0.45\,K\,km\,s$^{-1}$
      and are stepped by 0.3\,K\,km\,s$^{-1}$/0.15\,K\,km\,s$^{-1}$. 
    \end{enumerate}
    }
  \label{of-cb244}
\end{figure*}

\noindent{\bf CB~244} (Fig.~\ref{of-cb244}, left):
Apparent alignment of the polarization vectors between the dominant SE core and the NW source
parallel to the density enhancement between both sources.
If higher resolved polarization maps with a sufficient, statistically larger sample of polarization vectors
in this region will confirm this finding, it would support the hypothesis of matter infall along magnetic field lines
during the initial stage of molecular cloud collapse.

\noindent Close to the main (SE) source, however, the orientation of the polarization vectors
seems to change slightly towards a preferential direction parallel of the outflow axis
and therefore perpendicular to the orientation of the dust/gas ``bridge'' between both sources.
One has to consider that the orientations of the small number of polarization vectors in the SE core
show a large scatter. A higher resolved polarization map would be required in order 
to confirm the change of the polarization/magnetic field orientation towards the SE core.
This would also help to decide whether the magnetic field is aligned with the outflow
(and therefore oriented in the same direction as in the region of the density
enhancement between both sources) or if the outflow and/or other processes related to
the ongoing star formation process in the SE core cause(d) a change of the magnetic field orientation.

\noindent The core of CB~244 is slightly elongated in E-W direction and therefore 
-- in contrast to B~335 and CB~230 -- more aligned with
the mean polarization direction than perpendicular to it. However, if the change
of the polarization within the SE core is real, a similar scenario as described
for the other two globules is expected.

\noindent{\bf CB~54} (Fig.~\ref{of-cb244}, right):
This is a large Bok globule associated with the molecular cloud BBW~4 in a distance of about
1.1\,kpc (Brand \& Blitz~1993).
In contrast to B~335 and CB~230, the mean polarization direction is found to be almost 
{\em perpendicular} to the outflow axis. However, one has to consider that 
(a) the polarization pattern shows a large scatter of the orientation of
the individual polarization vectors and
(b) the much larger distance of the object (compared to the other investigated globules)
does not allow to resolve structures of comparable size, i.e., the orientation of the magnetic field
on the scale as measured in the case of the other globules is not known. The polarization measurements
trace a magnetic field structure which is extended much more further out into the surrounding
interstellar space and may therefore be much more representative for the interstellar
magnetic field structure in this region than for the local magnetic field structure of CB~54.

\noindent This globule is slightly elongated in S-E/N-W direction, almost parallel to the direction of
the mean polarization, which is in contrast to the Bok globules B~335 and CB~230.
 However, due to the much higher spatial distance of CB~54, we can not rule out that this 
globule consists of several substructures (smaller globule cores) which are not resolved
in our intensity maps. This possibility is supported by the finding that a small young
near-infrared stellar cluster which was probably born in CB~54 is projected against the
the dense core of this globule (Yun~1996, Launhardt~1996).

\begin{figure*}
  \resizebox{\hsize}{!}{\includegraphics{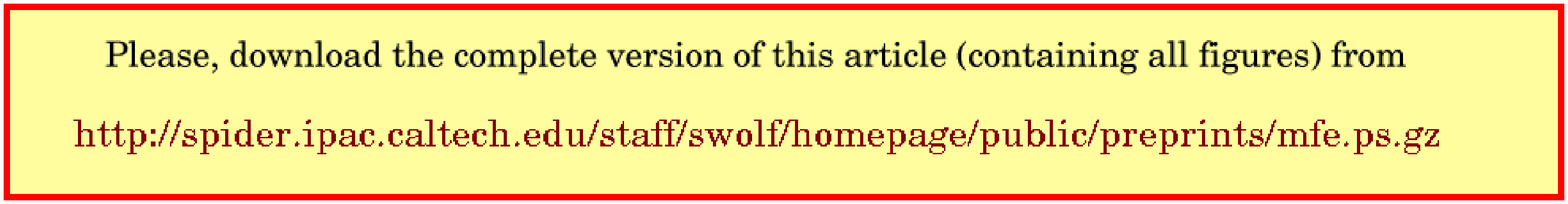}}
  \caption[]{
    {\bf Left:} SCUBA intensity map (850\,$\mu$m) of the Bok globule CB~26 with overlayed
    \begin{enumerate}
    \item {\em Iso-intensity contour lines} (thin dotted lines; 850\,$\mu$m; contour levels: 
      0.15\%, 0.30\%, 0.45\%, 0.60\%, 0.75\%, and 0.90\% of the maximum intensity $I_{\rm max}$),
    \item {\em Polarization pattern} (850\,$\mu$m; see Fig.~\ref{of-b335} for details).
    \end{enumerate}
    (from Paper~I).
    Furthermore, the mean direction of the polarization $\bar{\gamma}$ is shown.
    The range of $\bar{\gamma}\pm\sigma_{\bar{\gamma}}$ is marked.\\
    {\bf Right:}
    Central part of the Bok globule CB~26 (from Launhardt \& Sargent~2001).
    $J-H$ color map of the bipolar near-infrared reflection nebula (black contour lines: $K$ band emission).
    The white contour lines show the proto-planetary disk discovered by Launhardt \& Sargent~(2001;
    contour levels: 4, 11, 18, 25, and 32 mJy arcsec$^{-2}$; obtained with OVRO at 1.3\,mm - beam width: 
    0.58''$\times$0.39'').
  }
  \label{of-cb26}
\end{figure*}

\noindent{\bf CB~26} (Fig.~\ref{of-cb26}):
This small, slightly cometary-shaped Bok globule at a distance of about 140\,pc (Launhardt et al.~in prep.)
contains a small bipolar near-infrared nebula which is associated with strong submillimeter/millimeter continuum emission.
The star responsible for the reflection nebula (see Fig.~\ref{of-cb26}, right frame)
is deeply embedded and not seen even at 2.2\,$\mu$m.
No large-scale outflow was observed, but the edge-on circumstellar disk found by Launhardt \& Sargent~(2001)
as well as the bipolar structure perpendicular to the disk midplane suggest 
either the existence of an outflow in the plane of the sky (for which the velocity field 
could not be traced) or, at least, a very weak outflow.
The orientation of the mean polarization differs by about 35$^{\rm o}$ from the orientation of the disk.
As in the case of CB~54 we find the mean polarization direction (N-E/S-W)
to be slightly aligned perpendicular to the outflow direction.
The small number of polarization vectors does not allow to trace the expected randomly oriented polarization
vectors in the inner part of the globule.

\noindent The comparison with the results of an analytic investigation of the final states 
for a quasi-magnetostatic phase of the evolution of molecular cloud cores by ambipolar diffusion 
(based on a magnetized singular isothermal toroid model) by Li \& Shu~(1996) describes
the scenario probably found in the case of CB~26.
As Fig.~1 in their publication shows, an hourglass-shaped magnetic field structure is oriented
perpendicular to an already formed (proto)-circumstellar disk. Taking into account
the low resolution of our polarization maps, this magnetic field pattern translates
into an almost parallel polarization pattern with an orientation parallel to the disk and
perpendicular to the outflow, respectivly. Another prediction of their model is that at this stage
the magnetic field structure in the inner core would be very complex due to inflowing gas
colliding with the expanding outflows. In agreement with our observations, the polarization
vectors would be much more randomly oriented in the core. 
At this point we would like to remined the reader that this discussion is based on a statistically
very small number of polarization vectors only. A higher resolved polarization map is required
to confirm the conclusions drawn above.

\begin{deluxetable}{lrccl}
\tablecaption{Outflow and Magnetic Field orientations.\label{orient-concl}}
\tablehead{
\colhead{     }                         & 
\colhead{Mean Magnetic}                 &
\colhead{Mean Outflow}                  &
\colhead{Angle between outflow and}     \\
\colhead{     }                         & 
\colhead{field direction [$^{\rm o}$]}  &
\colhead{orientation [$^{\rm o}$]}      &
\colhead{Magnetic field orientation}
}
\startdata
CB~26  & $-65(\pm19)$ &  -29\tablenotemark{(a,e)} & 36\\
CB~54  & $ 22(\pm43)$ &   30\tablenotemark{(b)}   &  8\\
CB~230 & $-67(\pm30)$ &    0\tablenotemark{(b,c)} & 67\\ 
B~335  & $  3(\pm36)$ &  -80\tablenotemark{(d)}   & 83\\ 
CB~244 & $-22(\pm33)$ &   45\tablenotemark{(b)}   & 67\\
\enddata
\tablenotetext{(a)} {Launhardt \& Sargent~(2001)}
\tablenotetext{(b)} {Yun \& Clemens~(1994)}
\tablenotetext{(c)} {Launhardt~(2001)}
\tablenotetext{(d)} {Hirano et al.~(1988)}
\tablenotetext{(e)} {P.A.(disk): $60^{\rm o}\pm5^{\rm o}$ (Launhardt \& Sargent~2001)}
The error interval given in case of the magnetic field directions are equivalent	
to the standard deviation to the mean orientation angle of the polarization 
(see Tab.~\ref{obres} this paper and Tab.~2 in Paper~I).
\end{deluxetable}

We want to conclude the most interesting findings concerning the connections between the
magnetic field topology and Bok globule morphology of the globule cores (see also Tab.~\ref{orient-concl}):
\begin{enumerate}

\item   The globules B~335 and CB~230 show a slightly elongated shape which we
	assume in the following to be the projection of a prolate spheroid on the plane of the sky
	(in agreement with the theoretical and observational findings by Fiege \& Pudritz~2000a/c,
	Myers et al.~1991, and Ryden~1996 as discussed above).
	We exclude the CB~26 and CB~54 cores, which are also elongated, from this discussion.
	In contrast to the other sources, CB~26 is a remnant envelope around a young protostellar disk and CB~54
	is a large and massive globule which may contain multiple unresolved cores.

\item 	We find the direction of
	\begin{enumerate}
		\item The polarization vectors/magnetic field, 
		\item The elongation of the Bok globules, and
		\item The orientation of the outflows (in the case of CB~26: potential outflow direction) 
	\end{enumerate}
	to be not independent (=\,arbitrarily oriented) but related to one another.

\item   The direction of the outflows of B~335 and CB~230 is almost {\em perpendicular} to the orientation
	of the elongation direction of the globules, i.e., perpendicular to the potential projection
	of the symmetry axis on the plane of the sky.

\item 	In the case of B~335 and CB~230 the outflows are oriented almost {\em parallel} to the preferential direction
	of the linear polarization, i.e., perpendicular to the magnetic field (both as seen in projection	
	on the plane of the sky).

\item  	In the case of CB~54 and CB~26 the outflows are oriented slightly {\em perpendicular} 
	to the preferential direction of the linear polarization, i.e., parallel to the magnetic field.
	However, due to the large scattering of polarization directions measured in CB~54
	and the small number of polarization vectors and low spatial resolution in the case of CB~26,
	the statistical significance of these results is low.
	In the particular case of the globule CB~26, a spatially higher resolved polarization map would be required
	to confirm the apparent alignment between the direction of the potential outflow direction and the magnetic field.

\end{enumerate}
We exclude CB~244 from this overview
since we can hardly separate the magnetic field structure related to the density
enhancement reaching from the main (SE) source to the secondary (NW) source
from the magnetic field structure being dominant at the position of the main source alone.

We compared our polarization maps to those resulting from magneto-hydrodynamic (MHD) simulations of molecular clouds
performed by Padoan et al.~2001 (Fig.~5; see also Heitsch et al.~2001, Fig.~5 for comparison). 
In agreement with many of these simulations our observations of B~335, CB~230, and CB~26 show large-scale 
alignment of the polarization vectors with smooth changes of the orientation also on large scale only.
Similar to the simulated polarization pattern in Fig.~5d by Padoan et al.~(2001), the change
of the direction of the polarization vectors at the transition from the high-dense Bok globule
towards the lower-dense gas/dust ``bridge'' between the two cores in CB~244 can be seen.
Furthermore, our observations of B~335 and CB~230 show not only a decrease of the polarization
degree (see \S\ref{plvsi} for a detailed analysis), but also that the orientation of the polarization vectors 
does not remarkably change towards the dense cores. This supports the hypothesis that despite the
decrease of the grain alignment rate, the magnetic field structure in the cores of these objects
is not seriously disturbed, and thus still representing the primordial field.
In the case of CB~244 and CB~54 on the other hand, the orientation of the polarization vectors and therefore
the structure of the magnetic field in the cores is chaotic and we assume that it can not longer be accounted
for being representative for the primordial field. 
Thus, one might expect that CB~26 represents a more evolved protostellar systems than the globules B~335 and CB~230.
However, the data available so far, in particular the low-resolution, low-sampled polarization map of CB~26,
do not allow to place this assumption as a strong conclusion. A higher-resolved, better-sampled polarization map 
of CB~26 would help to confirm this hypothesis.

Spatially higher resolved polarization maps would simultaneously allow to test another theoretical
prediction about the protostellar evolution based on which the different orientations of the magnetic field relative 
to the outflow direction of B~335 and CB~230 on the one hand and CB~26 on the other hand could be explained:
Tomisaka~(1998) showed on the basis of MHD simulations of collapse-driven outflows in molecular cores that the direction of the magnetic
field lines and the disk plane decreases from 60$^{\rm o}$-70$^{\rm o}$ to 10$^{\rm o}$-30$^{\rm o}$
during the evolution of the outflow (Fig.~2a,b in his publication). 
At least on the large scale, our observations are in agreement with this scenario:
Since our polarization maps do not resolve structures with the size of a circumstellar disk, the magnetic field direction
as derived from the polarization maps, would change from an orientation {\em parallel to the disk midplane} to an orientation 
{\em perpendicular to the disk midplane} during the evolution of the protostar and its outflow in particular.

\section{Conclusions}\label{concl}

Using the Submillimeter Common User Bolometer Array (SCUBA) at the JCMT
we obtained 850\,$\mu$m polarization maps with a resolution of $9'\times9'$ of the Bok globules
B~335, CB~230, and CB~244. We find polarization degrees equally distributed in the range
$P_{\rm l}=0-14\,\%$. Using the formalism by Chandrasekhar \& Fermi~(1953) we derive
an estimate of the mean magnetic field strengths in these globules 
in the order of several hundred $\mu$G. These values are slightly higher than those
discussed in Paper~I based on another sample of Bok globules (B=20-100\,$\mu$G) 
but are still comparable to typical magnetic field strengths found in molecular clouds,
pre-protostellar cores, and other star-forming regions (see \S\ref{magfield} for references).
The magnetic fields derived here are higher because we based our calculations on a higher density closer to
the globule centers for the magnetic field estimates (see \S\ref{magfield}).

We find a similar correlation between the polarization degree and the intensity and therefore
the density of the emitting dust as measured in other star-forming cores (see \S\ref{plvsi}
for references). We verify the non-linear relation between these quantities 
first stated in Paper~I. However, the particular equation used to parameterize the decrease of the polarization
with increasing intensity (Eq.~\ref{nonlinfit}) is a first-order approximation only 
(see \S\ref{plvsi}, Fig.~\ref{allfits}). MHD simulations show qualitatively similar results (Padoan et al.~2001)
but a quantitative description of this phenomenon is still lacking.
Here, different grain (dis)alignment processes in the centers of Bok globules, such as discussed
in Paper~I will have to be considered in much more detail in these simulations
(see, e.g., Lazarian~1997 and Lazarian et al.~1997).

The main question we focussed on in this work is related to the search for correlations
between the structure of the magnetic field and particular features of Bok globules.
In addition to the globules B~335 and CB~230 we reconsider the globules
CB~26 and CB~54 (from Paper~I) as well. Because of the more complex structure
of the Bok globule CB~244, this object is only partly considered in this investigation 
(see \S\ref{magmorph}). Furthermore, CB~54 may represent an unresolved ensemble of Bok globules.
As common criteria to characterize the main spatial structure
of the Bok globules we take into account
(a) the orientation of the slightly elongated core (in case of B~335 and CB~230) and 
(b) the orientiation of the outflow (or the expected outflow in the case of CB~26).
Based on the theoretical studies by Fiege \& Pudritz~(2000a,c) and observational constraints
by Myers et al.~(1991) and Ryden~(1996) we assume that all these 4 globules have a prolate
shape rather than an oblate one.

We find that
{\em the outflows are oriented almost perpendicular to the symmetry axis of the globule cores}
in case of the globules B~335 and CB~230. The elongations found in case of CB~54 and CB~26
have a similar orientation, but the lower (absolute) spatial resolution of CB~54 and
the possible influence of a neighboring object of CB~26 may have changed the intrinsic globule shape.
The magnetic field, however, is aligned with the symmetry axis of the prolate cores 
in the case of the Bok globules B~335 and CB~230, while it is slightly aligned
with the outflow axis in the case of the Bok globules CB~26 and CB~54.
Since the symmetry axis of the core is expected to be aligned with the magnetic field
in order to explain the observed high abundance of prolate Bok globules,
we assume that the magnetic field structure found in the case of the globules B~335 and CB~230
represents the primordial magnetic field which is nearly undisturbed even in the innermost 
regions of these globule cores. The polarization decreases towards the centers of these cores
but the orientation of the linear polarization and therefore, prependicular to it, the
orientation of the magnetic field, is the same as on larger scales.
The polarization maps of these two objects allow to resolve structures with diameters
of about $2\times10^3$\,AU (B~335) and $4\times10^3$\,AU (CB~230). 
If the magnetic field structure in the circum-(proto)stellar environment differs strongly 
from the large-scale magnetic field direction, these pertubations must occur on much smaller
scales than given by the resolution of the polarization maps in order to allow the net polarization
to be aligned with the large-scale magnetic field structure. Furthermore, we cannot exclude the case
that the magnetic field structure is very complex on scales much smaller than
the resolution of the polarization maps. Then, the net polarization arising
from the inner core is likely to be negligible and all we measure is the small contribution
of polarized light from the foreground material.

Following the results of hydrodynamic simulations by Tomisaka~(1998) and 
theoretical investigations of the protostellar evolution by Li \& Shu~(1996), 
the orientation of the magnetic field relative to the outflow direction is not constant but changes during 
the evolution of the outflow/disk.
Taking into account the comparably low resolution of our polarization maps, this change would result in
\begin{enumerate}
\item A polarization pattern which is oriented parallel to the outflow
  ({\em mean magnetic field perpendicular to the outflow}) at the beginning of the outflow changing to
\item A negligible polarization in the range of the core due to polarization cancelation (averaging) effects
  at some point during the evolution of the outflow, and finally to
\item A polarization pattern which is oriented perpendicular to the outflow
  ({\em mean magnetic field parallel to the outflow}) at the late stage of the evolution of the outflow.
\end{enumerate}
Within this frame, our results for the Bok globules B~335 and CB~230 are consistent with an evolutionary stage somewhere 
between step 1 and 2.
Furthermore, the different relative orientation of the magnetic field relative to the outflow direction / circumstellar
disk orientation in case of CB~26 suggests that this object already reached evolutionary stage 3.
However, a better sampled, higher resolved (sub)millimeter polarization map is required in order to confirm this assumption.

We conclude that our observations suggest an evolution of the magnetic field of protostellar systems embedded in Bok globules.
Our findings do agree with theoretically predicted scenarios at this very early stage of stellar evolution, both in respect 
of the alignment of the magnetic field with the elongated globule as well as with the correlation between
the magnetic field direction relative to the outflow direction.

{}

\begin{acknowledgements}
The authors wish to acknowledge R.\ Tilanus for supporting the observations
and data reduction.
This research was supported by the DFG grant Ste\,605/10 within the program
``Physics of Star Formation'', and by the travel grant He~1935/22-1 of the DFG.
S.W.\ acknowledges support through the HST Grant GO\,9160, 
and through the NASA grant NAG5-11645.
JCMT is operated by the Joint Astronomical Centre on behalf of the UK Particle
Physics and Astronomy Research Council.
We wish to thank the anonymous Referee for helpful comments concering the discussion of the results.
\end{acknowledgements}

\end{document}